\documentclass[aps,pre,preprint,groupedaddress,showpacs,showkeys]{revtex4}
\usepackage{amsmath,amssymb}
\usepackage{graphicx}

\begin{document}

\title{Direct sample estimates of multidimensional quadratic statistical
functions: application to the anisotropic KPZ equation}

\author{Ivailo S. Atanasov and Oleg I. Yordanov}
\email[corresponding author:]{oiy@ie.bas.bg }
\thanks{the authors contributed equally to this paper}
\affiliation{Institute of Electronics, Bulgarian Academy of
Sciences, boul. ``Tzarigradsko Chousse\'e'' 72, 1784 Sofia,
Bulgaria}

\date{\today}

\begin{abstract}
We suggest a class of direct sample estimates for the two-point
quadratic statistical functions of multidimensional data, which
includes: estimates of the sample autocovariance function (AcF),
sample mean square increment (also, structure) function, and the
estimate of the power spectrum. The central estimate for the class
is the sample AcF, which is constructed as to represent the finite
Fourier transform of the periodogram estimate of the spectrum and
is positive semidefinite. The estimate explicitly account for the
anisotropy of the fields in all spatial directions and is
illustrated on two examples: the morphology of the Grab nebula and
the surface roughness generated as a solution of the anisotropic
Kardar-Parisi-Zhang equation.  We also provide an expression of
the covariance of the sample AcF in the case of data assumed to be
drawn from a two-dimensional Gaussian field with a known mean.

\end{abstract}


\maketitle

\section{\label{sec:intro}Introduction}

Despite the increased availability of multidimensional data, it
appears that not much attention has been devoted to the problem of
adequate and accurate direct estimation of simple quantities such
as two-point, quadratic statistical functions. The latter include
autocovariance (AcF) and autocorrelation functions, and the mean
square increment (also structure) function. Among the monographs
we are aware of, a good exception is \cite{Pr81}, where an
estimate of the two-dimensional AcF is discussed in some detail.
Specifically, for a given $N_1\times N_2$ matrix of real data
$f_{x_1x_2}$, $x_1=1, 2, \dots N_1$, $x_2=1, 2, \dots N_2$, the
autocovariance function is estimated in volume 2, chapter 9 of
\cite{Pr81} by
\begin{equation}\label{eq:S2dAcFs}
A_{\ast}(\vec{u}) =
        \displaystyle
        \frac{1}{N_1N_2}\sum_{x_1=1}^{N_1-u_1}\sum_{x_2=1}^{N_2-u_2}
        \hat{f}_{x_1, x_2}\;\hat{f}_{x_1+u_1, x_2+u_2},
\end{equation}
where $\vec{u}=(u_1, u_2)$, $\hat{f}_{x_1x_2}=
\left(f_{x_1x_2}-\bar{f}\right)$ with $\bar{f}$ denoting the
sample mean, $\bar{f}=(1/N_1N_2)
\sum_{x_1=1}^{N_1}\sum_{x_2=1}^{N_2}f_{x_1x_2}$. The
expression is valid for $0\leq u_1 \leq N_1-1$ and $0\leq u_2\leq
N_2-1$ and is extended to the third quadrant using the AcF
property of being even function. Formally, the latter replaces
$u_1$ and $u_2$ in (\ref{eq:S2dAcFs}) by their absolute values.
Note that this leaves $A_{\ast}(\vec{u})$ undefined in the second
and the fourth quadrants. We shall call this estimate ``the
standard'' 2d sample autocovariance function (SAcF) estimate.

For a homogeneous (stationary) random 2d field, $f(\vec{x})$, the
spectral representation theorem asserts that its AcF and spectral
density function (or simply the spectrum) are related by the
Fourier transform,
\begin{equation}\label{eq:four}
{\cal S}(\vec{k}) = \left(\frac{1}{2\pi}\right)^2
\sum_{x_1=-\infty}^{+\infty}\sum_{x_2=-\infty}^{+\infty} {\cal
A}(\vec{x})e^{i\vec{x}\cdot\vec{k}},
\end{equation}
where $\vec{k}=(k_1, k_2)$. We use calligraphic letters to
distinguish the ``true'' spectra and AcFs from their estimates.
The inverse transform of (\ref{eq:four}) reads:
\begin{equation}\label{eq:invfour}
{\cal A}(\vec{x})= \int_{-\pi}^{\pi}\int_{-\pi}^{\pi}{\cal
S}(\vec{k})e^{-i\vec{x}\cdot\vec{k}}d^2k.
\end{equation}

Equations~(\ref{eq:four}) and (\ref{eq:invfour}) apply to a random
field, which depends on discretely valued vector variable. It is
desirable to have estimates that are also related by a (discrete
and finite) Fourier transform. The simplest direct estimate of the
spectrum is provided by the periodogram
\begin{equation}\label{eq:Per}
I(\vec{k}) = \frac{1}{(2\pi)^2N_1N_2} \sum_{x_1,
y_1=1}^{N_1-1}\sum_{x_2, y_2=1}^{N_2-1}
\hat{f}_{x_1x_2}\hat{f}_{y_1y_2}
e^{-i\vec{k}\cdot(\vec{x}-\vec{y})}.
\end{equation}
The periodogram, referred to as ``raw'' (unsmoothed) estimate, is
a basis for variety of more sophisticated spectral estimates.
Hence, in accordance with (\ref{eq:invfour}), we would like to
have an AcF direct estimate, which for all $0\leq |u_1| \leq
N_1-1$ and $0\leq |u_2|\leq N_2-1$ satisfies
\begin{equation}\label{eq:Aest}
 A(\vec{u})=\int_{-\pi}^{\pi}\int_{-\pi}^{\pi}
 I(\vec{k})e^{-i\vec{k}\cdot\vec{u}}d^2k.
\end{equation}
The standard estimate, $A_{\ast}$,  satisfies (\ref{eq:Aest}), in
the domains where it is defined. However, it is easy to see that
if the validity of (\ref{eq:S2dAcFs}) is extended to the second
and the fourth quadrants, Eq. (\ref{eq:Aest}) does not hold. The
latter follows also from the developments presented in the next
section.

Another property of the theoretical AcF that \emph{must} be shared
by an estimate is the property of positive semidefiniteness. Let
$a_{x_1, x_2}$ be an arbitrary non-zero $\left(N_1\times
N_2\right)$ matrix of real numbers. Then the AcF of a 2d
stationary field satisfies
\begin{equation}\label{eq:psdef}
    \sum_{x_1, y_1=1}^{N_1} \sum_{x_2, y_2=1}^{N_2}
    a_{x_1, x_2}a_{y_1,y_2}{\cal A}\left(x_1-y_1,
    x_2-y_2\right)\geq0.
\end{equation}
It easy to see that $A_{\ast}$ does not satisfy (\ref{eq:psdef})
in the second and the fourth quadrants. We stress that the
inequality (\ref{eq:psdef}) does not bear theoretical importance
only; it ensures the positiveness of spectral estimates based on
an AcF estimate (indirect estimates) \cite{PW95}, see also
Eq.~(\ref{eq:Sest}) below.

In this paper we obtain a 2d AcF direct estimate, denoted
hereinafter by $A(\vec{u})$, which satisfies (\ref{eq:Aest}) in
all four quadrants, see the next section -- Eq. (\ref{eq:S2dAcF}).
The estimate is identical to $A_{\ast}(\vec{u})$ in the first and
the third quadrants but differs in the second and the fourth
quadrants. The estimate $A(\vec{u})$ is positive semidefinite and
leads to new estimates of both the mean square increment
(structure) function and the power spectrum. We briefly discuss
these estimates also in the second section and provide
generalization of (\ref{eq:S2dAcF}) for arbitrary dimension.

One of the most important advantages of the new estimate lays in
the fact that it is capable to capture the anisotropy of the data
in arbitrary spatial direction. The latter dictates the choice of
the illustrations and applications we consider here; yet another
application --- a study of an YBCO thin film morphology using
(\ref{eq:S2dAcF}) --- can be found in \cite{ADVBY06pha}. In the
first of the two applications, we study the AcF of images
representing the morphology of the Crab nebula. The random field
for these images is the light intensity in a recorded pixel. The
anisotropy in this case is determined either by the direction of
the expansion of the supernova ejecta or by the interaction of
synchrotron nebula with the ejecta \cite{Hester96apj,Hester98apj}.

In the third section we compute the covariance of $A(\vec{u})$ for
dimension $d=2$ and under the simplifying assumptions that the
observations are drawn from a Gaussian random field and are
already adjusted to have zero mean.  Our second application
involves the anisotropic Kardar-Parisi-Zhang (AKPZ) equation
\cite{Villain91jdpI,Wolf91prl,BS95}, considered in section four in
somewhat more detail. The equation pertains to growth of vicinal
surfaces and the anisotropy arises from the different rates of
growth along and across the average steps direction
\cite{Villain91jdpI}. We study how this anisotropy imprints on the
shape of the AcF by numerically solving the AKPZ equation on a
large lattice and then taking a smaller size images rotated on
various angles with respect to the axis of anisotropy. A summary
of our results and main conclusions are presented in the last
section.

\section{\label{sec:2dSAcF}Sample estimates of the multidimensional AcF}

In order to obtain a direct estimate of the 2d sample AcF that
corresponds to $I(\vec{k})$, we simply need to rearrange
(\ref{eq:Per}) to a 2d discrete Fourier transform.  We begin by
changing $y_1=x_1-u_1$ and reversing the order of $x_1$ and $u_1$
summation. This breaks up the sums with respect to $x_1, y_1$ to
two double sums: $\sum_{x_1=1}^{N_1}\sum_{y_1=1}^{N_1}
=\sum_{x_1=1}^{N_1}\sum_{u_1=x_1\!-\!1}^{x_1-N_1} =
\sum_{u_1=-(N_1\!-\!1)}^{0}\sum_{x_1=1}^{N_1+u_1} \;+\;
\sum_{u_1=1}^{N_1-1}\sum_{x_1=1+u_1}^{N_1}$. Next, we shift the
summation $x_1\mapsto x_1+u_1$ in the second term only obtaining
the intermediate result:
\begin{widetext}
\begin{eqnarray}\label{eq:P2}
\nonumber I(\vec{k}) = \frac{1}{(2\pi)^2N_1N_2}
\sum_{x_2,y_2=1}^{N_2} &\Bigg\{&
\sum_{u_1=-(N_1\!-\!1)}^{0}\sum_{x_1=1}^{N_1-|u_1|}
 \hat{f}_{x_1,x_2}\hat{f}_{x_1+|u_1|,y_2}
\;e^{-i[k_1u_1+k_2(x_2-y_2)]}
\\ \nonumber
&+& \sum_{u_1=1}^{N_1-1}\sum_{x_1=1}^{N_1-|u_1|}
\hat{f}_{x_1+|u_1|,x_2}\hat{f}_{x_1,y_2}
\;e^{-i[k_1u_1+k_2(x_2-y_2)]}\Bigg\}.
\end{eqnarray}
\end{widetext}
In analogous manner we deal with the $x_2$, and $y_2$ summations
obtaining four terms, which then can be combined into two pairs
arriving at the following 2d discrete Fourier representation of
the periodogram

\begin{equation}\label{eq:P3}
I(\vec{k}) = \frac{1}{(2\pi)^2}
\sum_{u_1=-N_1+1}^{N_1-1}\sum_{u_2=-N_2+1}^{N_2-1} A(\vec{u})
\;e^{-i\vec{k}\cdot\vec{u}}.
\end{equation}
In (\ref{eq:P3}), we introduced the function

\begin{equation}\label{eq:S2dAcF}
A(\vec{u}) = \left\{
    \begin{array}{l}
        \displaystyle
        \frac{1}{N_1N_2}\sum_{x_1=1}^{N_1-|u_1|}\sum_{x_2=1}^{N_2-|u_2|}
        \hat{f}_{x_1, x_2}\;\hat{f}_{x_1+|u_1|, x_2+|u_2|}
           \qquad  \mbox{for $u_1\cdot u_2\geq 0$}\\
        \displaystyle
        \frac{1}{N_1N_2}\sum_{x_1=1}^{N_1-|u_1|}\sum_{x_2=1}^{N_2-|u_2|}
        \hat{f}_{x_1, x_2+|u_2|}\;\hat{f}_{x_1+|u_1|, x_2}
        \qquad   \mbox{for $u_1\cdot u_2<
        0$},
    \end{array}
\right.
\end{equation}
which represents the new estimate of the sample AcF in dimension
two. It differs from the standard estimate (\ref{eq:S2dAcFs}) in
the second and fourth quadrants only. We stress that by the virtue
of its construction, it is (\ref{eq:S2dAcF}) but not the standard
estimate that is related to the periodogram by (\ref{eq:P3}).

A different way to obtain $A(\vec{u})$ is to substitute
$I(\vec{k})$ directly in Eq.~(\ref{eq:Aest}) and perform the
integration with respect to $\vec{k}$.  Then to use the obtained
pair of Kronecker symbols to carry out two of the summations
carefully accounting for the limits of summation. The latter
depends on the position of $\vec{u}$, which produces
(\ref{eq:S2dAcF}).

The estimate $A(\vec{u})$ is positive semidefinite function. This
follows from the way it was obtained but it is instructive to show
that the positive semidefiniteness virtually forces the form of
the SAcF, Eq.~(\ref{eq:S2dAcF}). To demonstrate this we use a 2d
generalization of a proof of semidefiniteness due to McLeod and
Jim\'{e}nez, see \cite{PW95}, chapter 6. Let $\varepsilon_{x_1,
x_2}$ be a 2d white noise with zero mean and variance $1/N_1N_2$.
Form a the field $g_{x_1, x_2}= \sum_{k=1}^{N_1}
\sum_{l=1}^{N_2}\hat{f}_{k, l} \varepsilon_{x_1-k, x_2-l}$; here
the centered values of the sample, $\hat{f}_{k, l}$, are
considered as coefficients of a moving average-like 2d field.
Since $g_{x_1, x_2}$ is stationary, its theoretical AcF must be
positive semidefinite. A straightforward calculation shows that
this function is given by (\ref{eq:S2dAcF}). One last remark
regarding (\ref{eq:S2dAcF}): it easy to see that the estimate
$A(\vec{u})$ can be computed numerically using the FFT-based
algorithm and the codes given in \cite{NR92}, chapters 12 and 13;
however, by first \emph{extending the data twice in both
dimensions and assuming that $f_{x_1,x_2}\equiv 0$ if either
$N_1\leq x_1\leq 2N_1$, or $N_2\leq x_2\leq 2N_2$} (double length
zero padding).

Estimate (\ref{eq:S2dAcF}) can be used as a basis for estimates of
both sample power spectrum and sample \emph{mean square increment
function}. One can obtain an estimate of the spectrum from
(\ref{eq:S2dAcF}) by employing it in a 2d generalization of
Grenander and Rosenblatt formula \cite{GR57},

\begin{equation}\label{eq:Sest}
S(\vec{k}) = \frac{1}{(2\pi)^2}\sum_{u_1=-N_1+1}^{N_1-1}
\sum_{u_2=-N_2+1}^{N_2-1} w(\vec{u})A(\vec{u})
e^{-i\vec{k}\cdot\vec{u}},
\end{equation}
where the function $w(\vec{u})$ is termed \emph{``lag window''}.
The general properties and specific examples of $w(\vec{u})$ can
be found in \cite{Pr81}; we note here that obviously it must be an
even function $w(-\vec{u})=w(\vec{u})$. The statistics of this
estimate will be presented elsewhere, note however that as in the
1d case, (\ref{eq:Aest}) provides a smoothed compared to the
``raw'' periodogram estimate.

The mean square increment (structure) function of $f(\vec{x})$,
${\cal B}(\vec{u})$, is defined \cite{Berry79jpa,Panchev71} by
${\cal B}(\vec{u})=E\left[\left(f(\vec{x}+\vec{u})-
f(\vec{x})\right)^2\right]$, where $E$ denotes the ensemble
averaging, and is related to the AcF by ${\cal
B}(\vec{u})=2\left({\cal A}(0)-{\cal A}(\vec{u})\right)$,
e.g.~\cite{YA02epj}; for an estimate that corresponds to the
standard estimate of ${\cal A}(\vec{u})$ see
\cite{Russ94,TRA99wear}. An estimate corresponding to
(\ref{eq:S2dAcF}) should be modified in the second and the fourth
quadrants as:
\begin{equation}\label{eq:Best}
 B(\vec{u})=\frac{1}{N_1N_2}\sum_{x_1=1}^{N_1-|u_1|}\sum_{x_2=1}^{N_2-|u_2|}
 \left(f_{x_1, x_2+|u_2|}-f_{x_1+|u_1|, x_2}\right)^2,
 \qquad   \mbox{for $u_1\cdot u_2<0$}.
\end{equation}
In addition of being a positive function, we remark that
$B(\vec{u})$ does not involve the sample mean of $f(\vec{x})$ and
thus is free from a source of bias brought up by $\overline{f}$
\cite{BJ76,PW95}. Modifications analogous to (\ref{eq:Best}) are
apparently due for the estimates of the generalized structure
functions used to infer multifractal scaling, see
Refs.~\cite{BV91pra,Krug94prl}.

Estimate (\ref{eq:S2dAcF}) can readily be generalized to arbitrary
dimension $d$. To shorten the notations, consider a multi-index
with $d$ components $p=(p_1, p_1, \dotsc p_d)$ each taking a value
of either $0$ or $1$. Let $p_k=0$ indicates $x_k$, whereas $p_k=1$
indicates $x_k+|u_k|$, $k=1, 2, \dotsc d$, and let $\overline{p}$
designate the multi-index whose components are all different from
the components of $p$; e.g. if $p=(011)$, then
$\overline{p}=(100)$. Then the $d$-dimensional SAcF estimate can
be expressed as
\begin{equation}\label{eq:SdAcF}
A_p(\vec{u}) =\frac{1}{N_1N_2\dotsc N_d}\sum_{x_1=1}^{N_1-|u_1|}
\dotsb \sum_{x_d=1}^{N_d-|u_d|} \hat{f}_p \hat{f}_{\overline{p}},
\end{equation}
where $p=(0, 0,\dotsc, 0)$ pertains to the first and the
$2^{d-1}+1$-th hyperquadrants, $p=(0, 0,\dotsc, 1)$ to the second
and $2^{d-1}+2$-th hyperquadrants, and so on. In general, the SAcF
in hyperquadrants $k$ and $2^{d-1}+k$ is expressed by
(\ref{eq:SdAcF}) with a multi-index p, which is the binary
representation of number $k-1$.

\begin{figure}
\hphantom{}\hspace{0.08\textwidth}
\includegraphics[width=0.9\columnwidth]{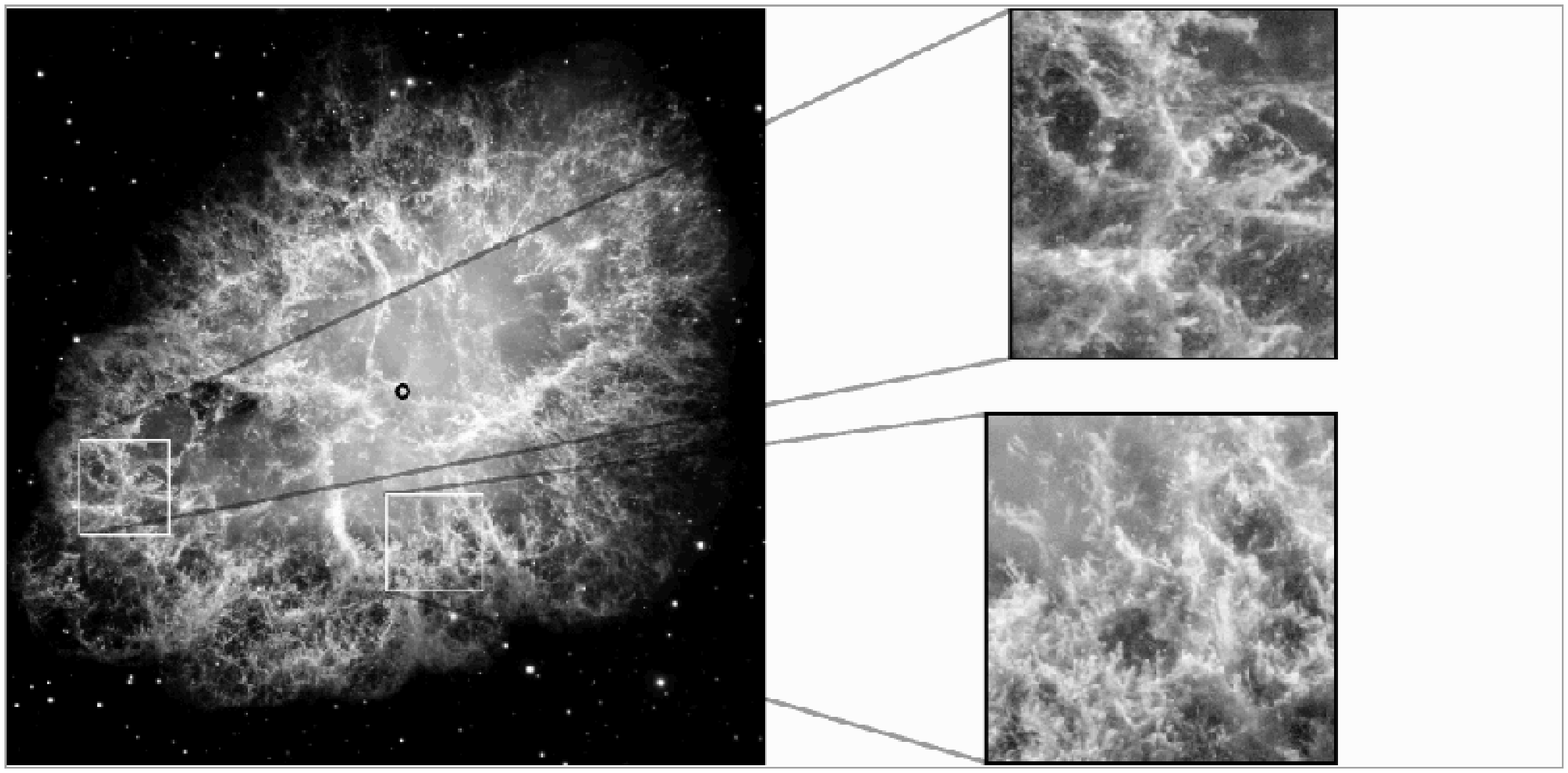}\\[0.5\baselineskip]%
\includegraphics[width=0.45\columnwidth]{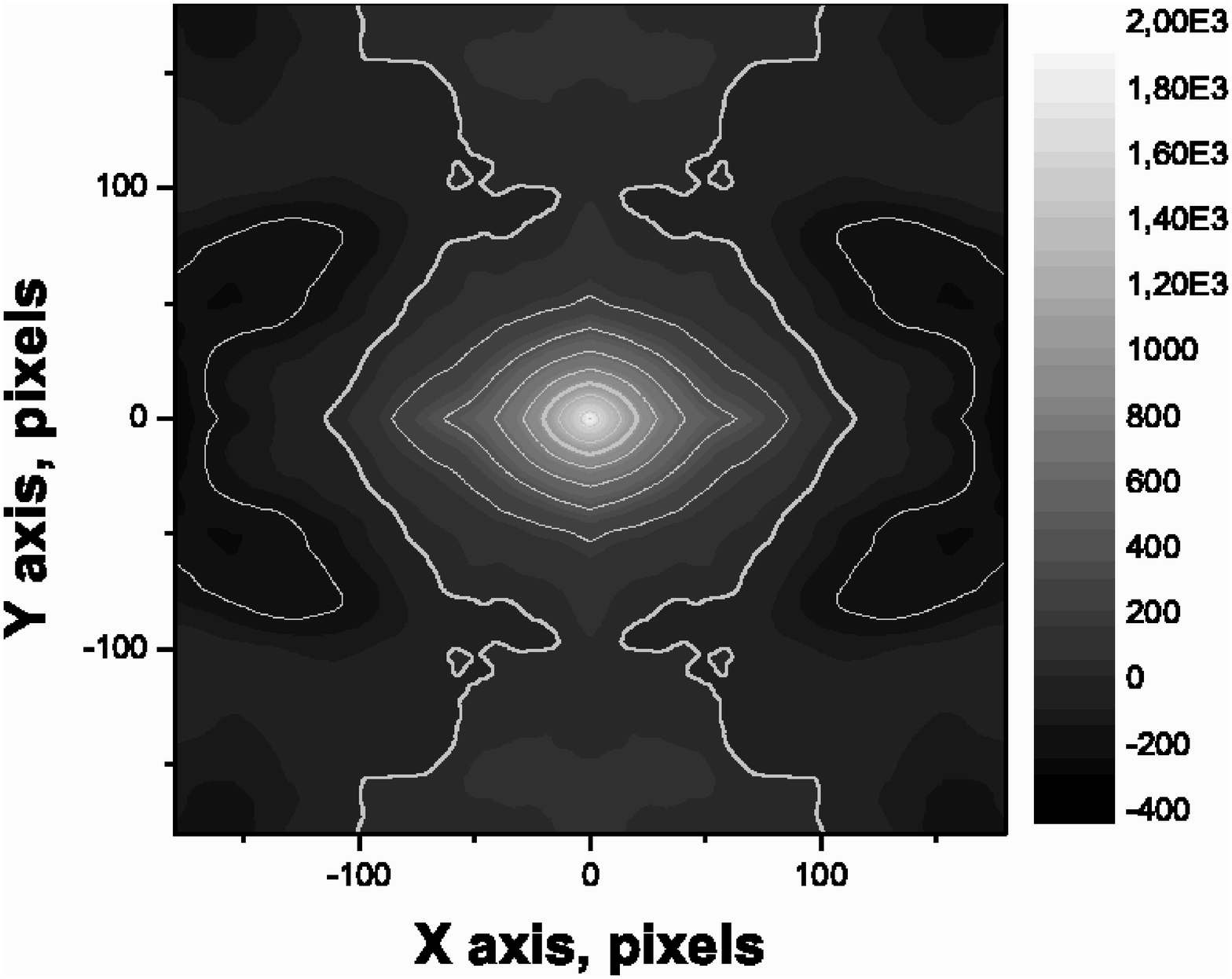}
\hfill
\includegraphics[width=0.45\columnwidth]{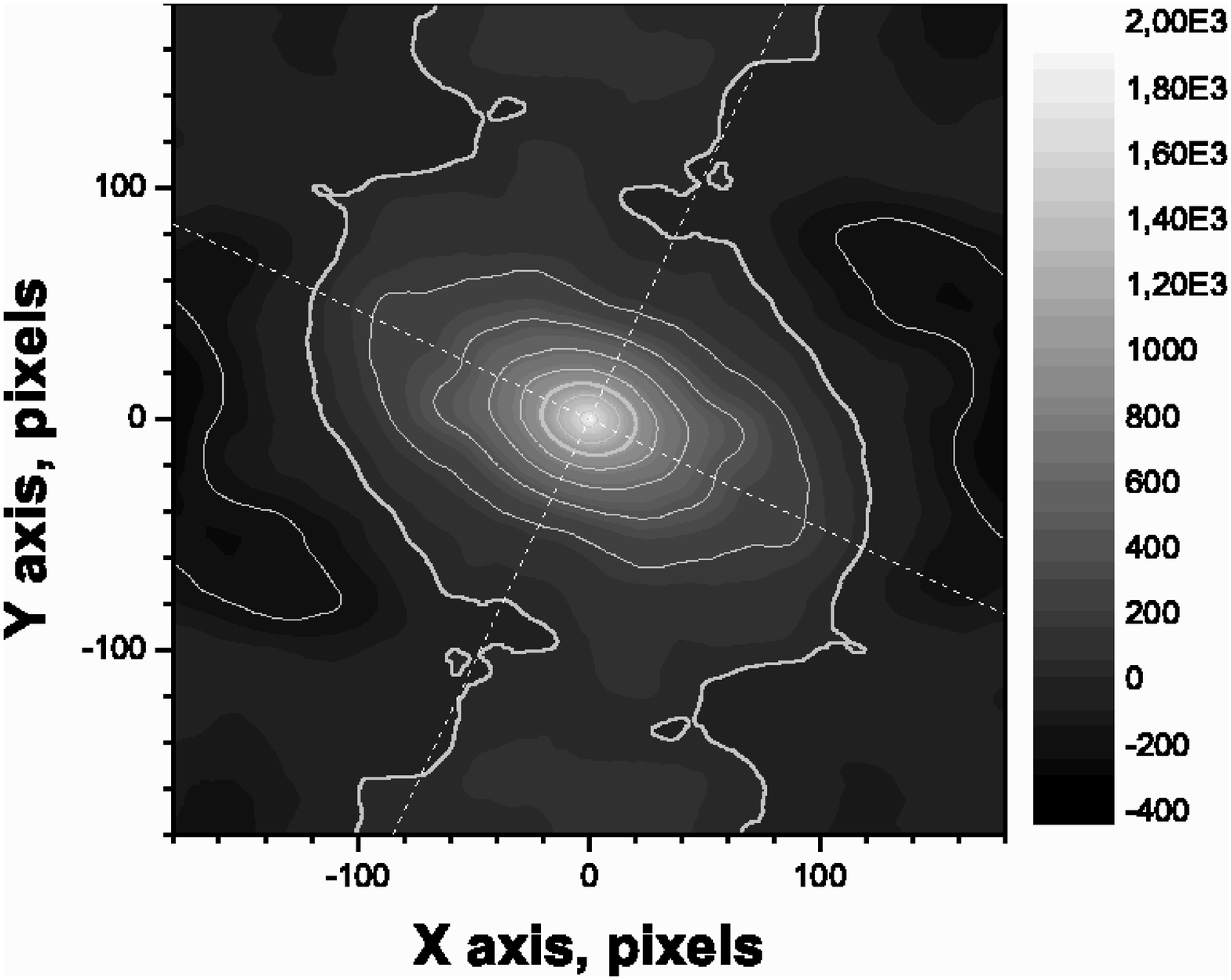} \\[0.5\baselineskip]%
\includegraphics[width=0.45\columnwidth]{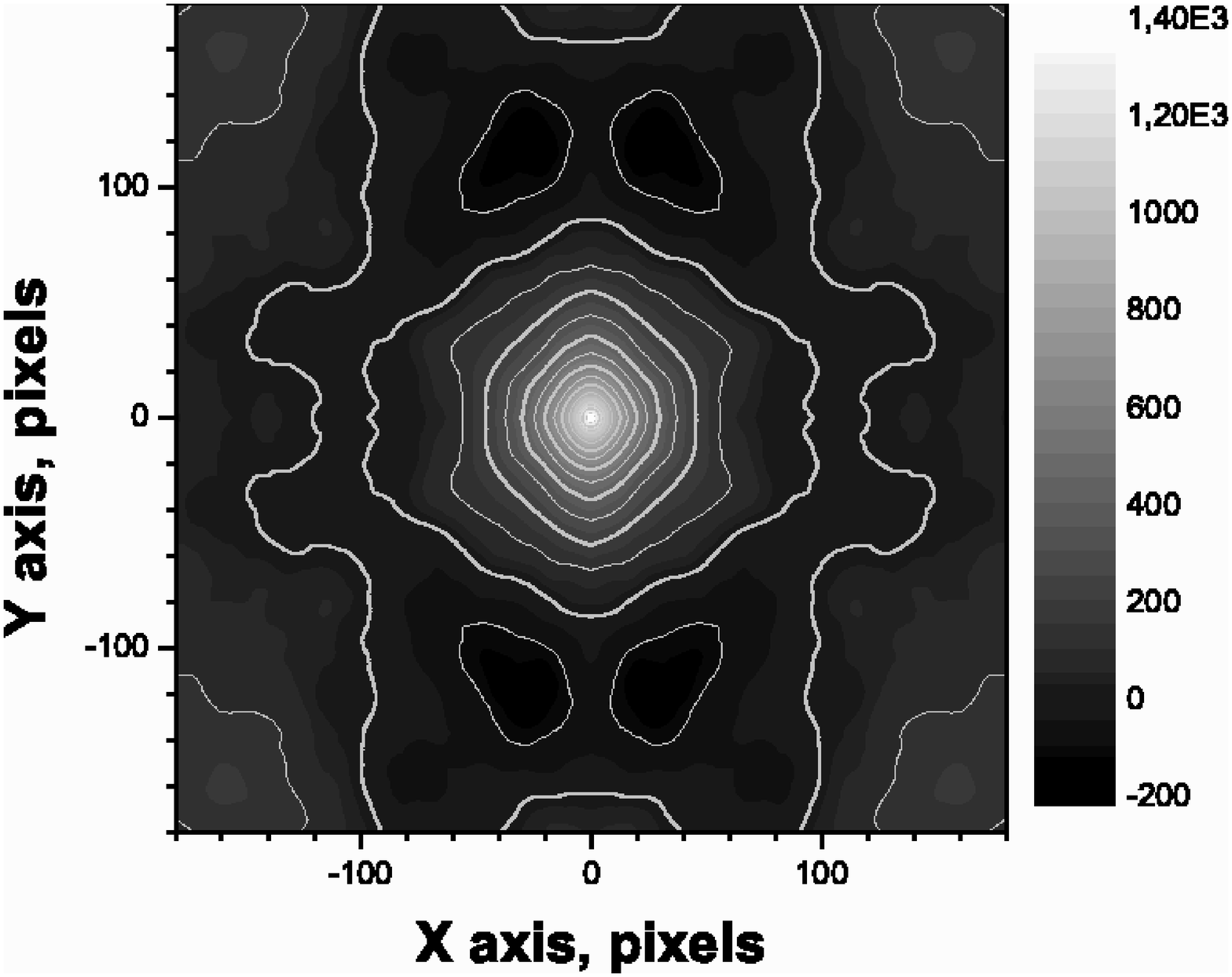}
\hfill
\includegraphics[width=0.45\columnwidth]{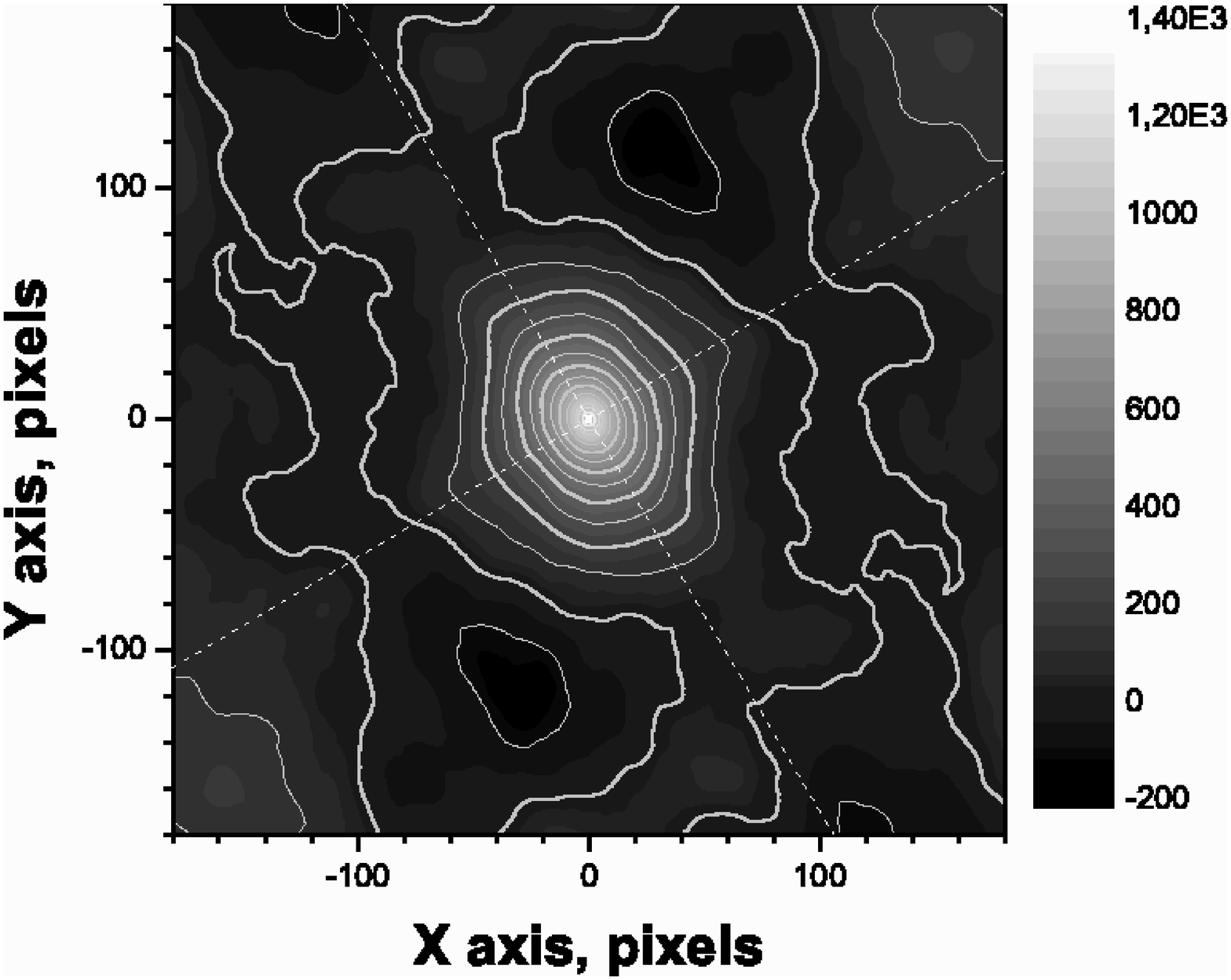}\\[0.5\baselineskip]%
\begin{flushleft}
\raisebox{1.25\textwidth}[0pt][0pt]{\large\bf (a)}
\hspace{0.52\textwidth}
\raisebox{1.25\textwidth}[0pt][0pt]{\large\bf (b)}\\[-1\baselineskip]
\raisebox{0.77\textwidth}[0pt][0pt]{\large\bf (c)}
\hspace{0.46\textwidth}
\raisebox{0.77\textwidth}[0pt][0pt]{\large\bf (d)}\\[-1\baselineskip]
\raisebox{0.38\textwidth}[0pt][0pt]{\large\bf (e)}
\hspace{0.46\textwidth}
\raisebox{0.38\textwidth}[0pt][0pt]{\large\bf (f)}
\end{flushleft}
\vspace{-2.5\baselineskip} \caption{\label{f:1} Standard AcF
estimate and estimate (\ref{eq:S2dAcF})  for two regions of the
Crab nebula. Panel (a) shows the location of the regions; the
circle indicates the position of the pulsar. Panel (b) shows the
two regions zoomed by an identical factor. Panels (c) and (d) show
the standard AcF estimate and the estimate (\ref{eq:S2dAcF}),
respectively for the upper image in (b). Panels (e) and (f) -- the
same the bottom image in (b). The morphology principle axes of
anisotropy are drawn in plates (d) and (f).}
\end{figure}

To illustrate the difference between the standard and the estimate
(\ref{eq:S2dAcF}), we provide plots of both estimates for pair of
images representing regions of the Crab nebula. The images were
selected from a color image taken from the \emph{Hubble Space
Telescope} (HST) \cite{wwwcrab}. The color image was created as a
weighted sum of three narrowband filters centered at 5012 \AA,
6306 \AA\ and 6732 \AA\ and comprises 24 individual Wide Field and
Planetary Camera 2 exposures~\cite{Hester96apj}. We converted the
color image into gray scale (in the range of 0-256) image, i.e.
$f_{x,y}$ in this case is proportional to the light intensity per
pixel and represents the morphology of the nebula; we added a
circle indicating the position of the Crab pulsar. The first of
the selected images, shown in Fig.~\ref{f:1}(b) top panel, is
located close to outer rim of the nebula dominated by the
expanding ejecta  \cite{Hester96apj,Hester98apj}. The coordinates
of its right bottom corner are, RA: $\alpha_{2000}=$5:34:40.5  and
Dec: $\delta_{2000}=$21:59:39.1. The image extends
(46.2$\times$49.7) arcseconds corresponding to $(N_1=453) \times
(N_2=487)$ pixels. By just inspecting the image the anisotropy is
not easily recognizable, however, due to the outward expansion of
the supernova remnant an anisotropy roughly across the radial
direction (direction to the pulsar) should present in the
morphology.

The second image has coordinates: $\alpha_{2000}=$5:34:29.0,
$\delta_{2000}=$ 21:59:10.0 and extends 51.8$\times$51.8
arcseconds, ($504 \times 504$) pixels --- Fig.~\ref{f:1}(b),
bottom panel. It is from a region where the synchrotron nebula
(upper left sector of the image) interacts with the denser ejecta
creating ``filaments''. The latter are attributed to a magnetic
Rayleigh-Taylor (R-T) instabilities \cite{Hester96apj}. The major
axis of anisotropy in this region should, in general, be expected
along rather then across the direction to the pulsar.

An overall linear (planar) trend,
$g=s_{\alpha}\alpha+s_{\delta}\delta$ with origins of coordinate
systems at the left-upper corners of the images, is removed before
both SAcF estimates were computed. Carrying out this procedure is
important since the trend by itself produces anisotropy in the
SAcF. In the case of images presented in Fig.~\ref{f:1}(b) the
linear trend is rather small; the estimated slopes (in units
grayscale/pixel) are $s_{\text{DEC}}=-0.027$,
$s_{\text{RA}}=0.0071$, and $s_{\text{DEC}}=0.101$,
$s_{\text{RA}}= 0.031$, for first and second images, respectively.
The standard and the estimate (\ref{eq:S2dAcF}) for the first
image are presented as gray scale plots with superimposed level
contours in Fig.~\ref{f:1}(c) and Fig.~\ref{f:1}(d), respectively.
For the sake of completeness of the plot the standard AcF is
extended to the second and the fourth quadrants, hence the
specific ``rose'' appearance of the AcF. The anisotropy of the
supernova remnant structures in this part of the nebula is clearly
recognized from the plot of $A(\vec{u})$ with the major axis of
anisotropy having an angle of about $\vartheta\approx -25^{\circ}$
with respect to the horizontal axis. This angle should be
interpreted as the average front of the local expansion, refer to
Fig.~\ref{f:1}(a). Another quantity that characterizes the
anisotropy is the \emph{aspect ratio}, $\gamma$, the ratio between
the characteristic sizes of the nebula structures along the minor
and majors axes of anisotropy. The latter sizes are defined by the
correlation lengths of SAcF in the respective directions. We
evaluate these lengths crudely by assuming, somewhat arbitrary,
that the 1d principal profiles of AcF are represented by random
processes with finite domain (band-limited) spectra. Using the
expressions for the correlation length obtained for this class of
random processes in \cite{YN94pre,YN97phd}, we infer
$\gamma\approx 0.57$.

For the second region, the major axis of the anisotropy is at
angle of $\vartheta\approx -59.5^{\circ}$ with respect to the
horizontal axis, which as should be expected is roughly in
direction to the pulsar, refer to Fig.~\ref{f:1}(a) and
Fig.~\ref{f:1}(f).  The aspect ratio in this case is
$\gamma\approx 0.875$.

\section{\label{sec:cvSAcF} Covariances of the 2d sample autocovariance function}

\begin{figure}
\includegraphics[width=0.9\columnwidth]{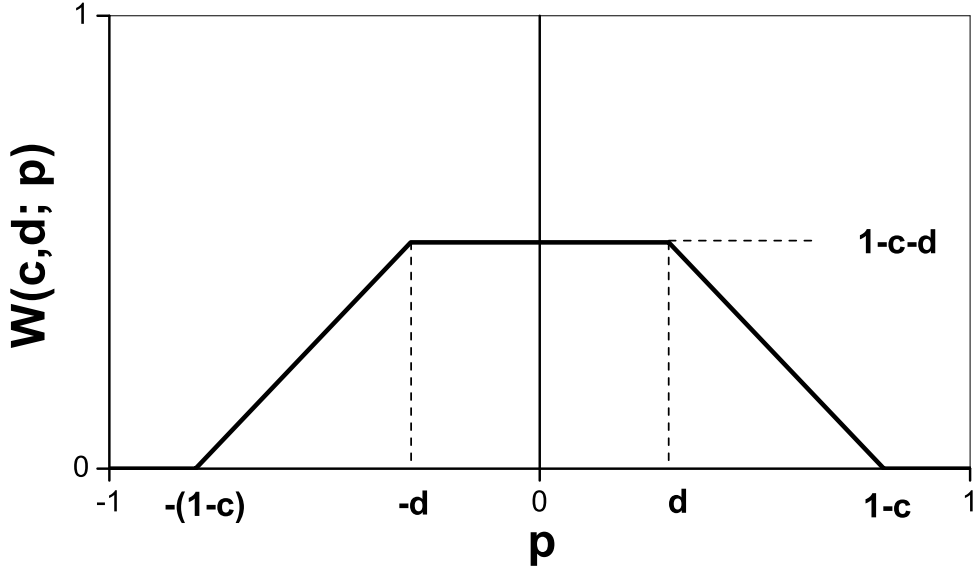}
\caption{\label{f:win} Graph of the window function $W(c,d; p)$,
Eq.~(\ref{eq:2dsacf-W}), used in the expression of the sample AcF
covariances.}
\end{figure}

In this section we obtain expression for covariances of 2d SAcF
 --- the estimate (\ref{eq:S2dAcF}), evaluated at two points
$\vec{u}=(u_1, u_2)$ and $\vec{v}=(v_1,v_2)$:
\begin{equation}
\label{eq:S2dAcF-cov}
\text{cov}\big[A(\vec{u}),A(\vec{v})\big] = E\big[ A(\vec{u})A(\vec{v})\big]
- E\big[A(\vec{u})\big]E\big[A(\vec{v})\big].
\end{equation}
The covariance has both theoretical as well as practical
importance for determining the confidence intervals in the AcF
estimate. The expression will be derived under the simplifying
assumptions that $f_{x_1,x_2}$ is a Gaussian random field with
zero mean (or that the mean is known and subtracted). It is
immediately seen that irrespective to which quadrant $\vec{u}$
belongs,

\begin{equation}
E\big[ A(\vec{u})\big] = \left(1 -
\frac{|u_1|}{N_1}\right) \left(1 - \frac{|u_2|}{N_2}\right) {\cal
A}(\vec{u}), \label{eq:S2dAcF-mean}
\end{equation}
where as before ${\cal A}(\vec{u})$ denotes the true
autocovariance function of $f(\vec{x})$.
Eq.~(\ref{eq:S2dAcF-mean}) is identical to the ensemble average of
the standard estimate and shows that (\ref{eq:S2dAcF}) is a biased
estimate. The bias, however, is small for large samples especially
for small $|\vec{u}|$. We turn now to the first term in
(\ref{eq:S2dAcF-cov}). Reckoning with $A(-\vec{u})=A(\vec{u})$ and
the symmetry under the exchange $\vec{u}\leftrightarrow\vec{v}$,
we need to consider three different cases only: (i) $\vec{u} \in$
I quadrant, $\vec{v} \in$ I quadrant; (ii) $\vec{u} \in$ II
quadrant, $\vec{v} \in$ II quadrant; and (iii) $\vec{u} \in$ I
quadrant, $\vec{v} \in$ II quadrant. The calculations in all three
cases are closely similar; below we illustrate them for the case
(ii) only. Inserting the pertinent for this case AcF expressions
from (\ref{eq:S2dAcF}) and using that for a Gaussian field the
four-point function can be expressed as combinations of products
of two two-point AcFs we have

\begin{eqnarray}
\label{eq:2dsacf-cov2}
  E\big[ A(\vec{u})A(\vec{v})\big] &=&
  \frac{1}{N_1^2N_2^2}
\sum_{x_1=1}^{N_1-|u_1|}\sum_{x_2=1}^{N_2-|u_2|}
  \sum_{y_1=1}^{N_1-|v_1|}\sum_{y_2=1}^{N_2-|v_2|}
 \Big[ {\cal A}(|u_1|,-|u_2|){\cal  A}(|v_1|,-|v_2|) \nonumber
\\
  &\quad& + {\cal  A}(y_1\!-\!x_1,y_2\!-\!x_2\!+\!|v_2|\!-\!|u_2|)
      {\cal  A}(y_1\!-\!x_1\!+\!|v_1|\!-\!|u_1|,y_2\!-\!x_2)
\\
\nonumber
 &\quad& + {\cal  A}(y_1\!-\!x_1\!+\!|v_1|,y_2\!-\!x_2\!-\!|u_2|)
      {\cal  A}(y_1\!-\!x_1\!-\!|u_1|,y_2\!-\!x_2\!+\!|v_2|)
  \Big].
\end{eqnarray}
The first of these terms does not depends of the summation indexes
and cancels out with the second term in (\ref{eq:S2dAcF-cov})
exactly, refer to (\ref{eq:S2dAcF-mean}). In the remaining two
terms we perform the indicated change $\vec{p}=\vec{y}-\vec{x}$
and again reverse the order of the summation. This allows to carry
out the summations with respect to both $x_1$ and $x_2$
explicitly, noting in the process that we need to distinguish the
case $|u_1|>|v_1|$ from the case $|v_1|>|u_1|$. The result is a
product of two trapezium shaped functions, which involve two
parameters, $c$ and $d$,
\begin{equation}\label{eq:2dsacf-W}
W(c,d;p) := \left\{
    \begin{array}{rrcl}
        0, & \quad \!&\!p\!&\! \le -(1-c)\\
        1-c+p, & \quad -(1-c) \le \!&\!p\!&\! \le -d\\
        1-c-d, & \quad -d \le \!&\!p\!&\! \le d\\
        1-c-p, & \quad d \le \!&\!p\!&\! \le 1-c\\
        0, & \quad 1-c \le \!&\!p\!&\!,
    \end{array}
\right.
\end{equation}
see also Fig.~\ref{f:win}. The parameters $c$ and $d$ are subject
to the conditions $0 \le d \le 1-c \le 1$.

Next, introducing $\vec{a} := (\vec{u}+\vec{v})/2$; $\;\vec{b} :=
(\vec{u}-\vec{v})/2$ and shifting the summation indexes $p_1$ and
$p_2$ simultaneously according to $\vec{p} =
(q_1-u_1-N_1,q_2+u_2-N_2)$, we arrive at the following expression
for the covariance valid when both $\vec{u}$ and $\vec{v}$ are in
the second quadrant:

\begin{widetext}
\begin{eqnarray}
\label{eq:S2dAcF-cov2} \nonumber
\mbox{cov}\Big[A(\vec{u}),A(\vec{v})\Big] &=& \frac{1}{N_1N_2}
\sum_{q_1=1}^{2N_1+2a_1-1}\; \sum_{q_2=1}^{2N_2-2a_2-1}
W\left(-\frac{a_1}{N_1},\frac{|b_1|}{N_1};\frac{r_1}{N_1}\right)
W\left(\frac{a_2}{N_2},\frac{|b_2|}{N_2};\frac{r_2}{N_2}\right)\\
&\quad& \times \Big[ {\cal A}(\vec{r}+\vec{a})\; {\cal
A}(\vec{r}-\vec{a}) + {\cal A}(\vec{r}+\vec{b})\; {\cal
A}(\vec{r}-\vec{b}) \Big],
\end{eqnarray}
\end{widetext}
where $\vec{r}=(q_1-a_1-N_1,q_2+a_2-N_2)$ has been introduced.

Similar expressions are obtained in cases (i) and (iii) above.
Finally, if we define
$\widetilde{a}_{1,2}:=(|u_{1,2}|+|v_{1,2}|)/2$ and
$\widetilde{b}_{1,2}:=(|u_{1,2}|-|v_{1,2}|)/2$, all three cases
can be combined into the following compact form:

\begin{widetext}
\begin{eqnarray}
\label{eq:S2dAcF-cov-f} \nonumber
\mbox{cov}\Big[A(\vec{u}),A(\vec{v})\Big] &=& \frac{1}{N_1N_2}
\sum_{q_1=1}^{2N_1-2\widetilde{a}_1-1}\;
\sum_{q_2=1}^{2N_2-2\widetilde{a}_2-1}
W\left(\frac{\widetilde{a}_1}{N_1},\frac{|\widetilde{b}_1|}{N_1};\frac{r_1}{N_1}\right)
W\left(\frac{\widetilde{a}_2}{N_2},\frac{|\widetilde{b}_2|}{N_2};\frac{r_2}{N_2}\right)\\
&\quad& \times \Big[ {\cal A}(\vec{r}+\vec{a})\; {\cal
A}(\vec{r}-\vec{a}) + {\cal A}(\vec{r}+\vec{b})\; {\cal
A}(\vec{r}-\vec{b}) \Big],
\end{eqnarray}
\end{widetext}
with the general definition $\vec{r}=(q_1+\widetilde{a}_1-N_1,
q_2+\widetilde{a}_2-N_2)$. This expression is valid for arbitrary
positions of vectors $\vec{u}$ and $\vec{v}$.

The important for the practice variances of the SAcF, can be
obtained from the expressions of the covariance. Setting
$\vec{v}=\vec{u}$ in (\ref{eq:S2dAcF-cov-f})   we have
\begin{widetext}
\begin{eqnarray}
\label{eq:S2dAcF-var}
\nonumber
  \mbox{var}\left[A(\vec{u})\right] &=& \frac{1}{N_1N_2}
  \sum_{p_1=-N_1\!+\!|u_1|\!+\!1}^{N_1\!-\!|u_1|\!-\!1}\;
  \sum_{p_2=-N_2\!+\!|u_2|\!+\!1}^{N_2\!-\!|u_2|\!-\!1}
  \left(1-\frac{|u_1|+|p_1|}{N_1}\right)
  \left(1-\frac{|u_2|+|p_2|}{N_2}\right)\\
  &\times& \Big[ \mathcal{A}^2\big(p_1,p_2\big) \;+\;
  \mathcal{A}\big(p_1\!+\!|u_1|,p_2\!+\!|u_2|\big)\;
  \mathcal{A}\big(p_1\!-\!|u_1|,p_2\!-\!|u_2|\big) \Big].
\end{eqnarray}
\end{widetext}
Note that we went back from summation with respect to $q_1$ and
$q_2$ to summation with respect to $p_1$ and $p_2$, which results
in symmetric about zero limits of summation.


\section{\label{sec:AKPZ}Application to the anisotropic KPZ equation}

The anisotropic Kardar-Parisi-Zhang (AKPZ) equation has been
introduced in an attempt to model the growth on a vicinal
substrates \cite{Villain91jdpI}. Adatoms that migrate towards the
steps and attach to them, have lower probability to desorb
compared to those migrating parallel to the steps. This
effectively induces different rates of growth along and across the
steps and violates the rotational symmetry of the KPZ growth
process \cite{KPZ86prl}. Let $h=h(\vec{x}, t)$ be the height of
the growing surface at point $\vec{x}=(x, y)$ and time $t$. If one
chooses the $x$-coordinate along the direction of the steps, the
AKPZ equation takes the form:
\begin{equation}\label{eq:akpz0}
   \partial_th=\nu_x\partial_x^2h+\nu_y\partial_y^2h+
   \frac{\lambda_x}{2}\left(\partial_xh\right)^2+
   \frac{\lambda_y}{2}\left(\partial_yh\right)^2 + \eta,
\end{equation}
see also \cite{BS95}.  In this equation: $\nu_x$ and  $\nu_y$ are
coefficients of the curvature terms associated with desorption,
$\lambda_x$ and $\lambda_y$ are coefficients related to growth
rates normal the surface, and $\eta=\eta(x, y, t)$ is a Gaussian
white noise, $E[\eta(x, y, t)\eta(x', y', t')] =
2D\delta(x-x')\delta(y-y')\delta(t-t')$. The equation has been
studied by D.~E.~Wolf using one-loop, renormalization-group (RG)
approximation \cite{Wolf91prl}. Some of the obtained results have
later been confirmed by numerical simulations \cite{H-HA92pra}. To
recap what will be needed here, let $r_{\nu}=\nu_y/\nu_x$ and
$r_{\lambda}=\lambda_y/\lambda_x$ and let both $\lambda_x$ and
$\lambda_y$ be positive. In this case the AKPZ surface grows with
an exponent identical to the surface generated by the isotropic
KPZ equation, referred to as algebraic roughness. As the
morphology evolves, the nonlinear parameters $\lambda_x$ and
$\lambda_y$, as in the case of isotropic KPZ are not renormalized,
whereas $\nu_x$ and $\nu_y$ take effective values such that
$r_{\nu}=r_{\lambda}$ (a fixed point to the dynamical
renormalization flow equations) \cite{Wolf91prl}. This means that
in this case the anisotropy of the surface is of the simplest kind
-- elliptical anisotropy -- and therefore might be taken as a
benchmark for testing statistical methods characterizing
anisotropy.

The numerical simulations were carried out using the Amar and
Family numerical scheme~\cite{AF90pra,CT89prb,MKW91pha}, which
broadly speaking includes rescaling of the equation and employing
the standard discretization for the derivatives. Two comments are
in order. First, in contrast to \cite{H-HA92pra}, we choose
rescaling that leaves the equation manifestly anisotropic: $x
\mapsto \sqrt{\nu_x}x$, $y \mapsto \sqrt{\nu_x}y$, $h \mapsto
(2\nu_x/\lambda_x)h$, and $\eta \mapsto \sqrt{2D/\nu_x}\eta$.
Second, the discrete analog does not adequately represent the
continuous AKPZ equation \cite{LS98pre1}, however, since more
accurate difference scheme are not known for dimensions higher
than one \cite{LS98pre2,Buceta05pre}, we employ here the standard
discretization. What is more important within the scope of this
study, the discrete equation inherits the elliptical anisotropy of
the original AKPZ equation.

\begin{figure}
\includegraphics[width=0.8\columnwidth]{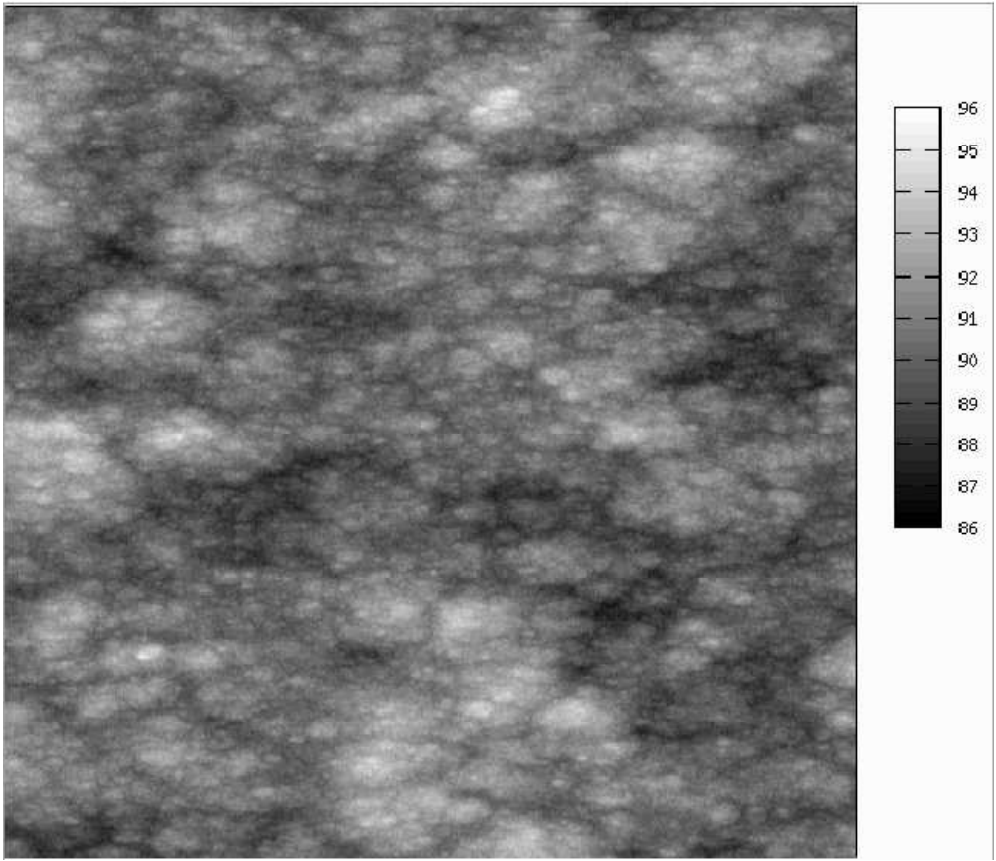}
\caption{\label{f:akpzsurf} Gray scale image of a morphology
obtained by numerically solving the anisotropic KPZ equation. The
image represents the central (512$\times$512) part  from the
entire (1024$\times$1024) simulated surface. The conditions at
which the simulation is carried out are discussed in the text. }
\end{figure}

The elliptical anisotropy can be discerned even by visually
inspecting of the simulated morphology, see Fig.~\ref{f:akpzsurf}.
The picture represents AKPZ surface generated with parameters
$\nu_x=1$, $\nu_y=0.3$,  $\lambda_x=10$, $\lambda_y=3$, and
$D=0.2$; (to skip the ``transient'' time for the system to come to
the RG fixed point, we have chosen $r_{\nu}=r_{\lambda}=0.3$ at
the outset). The simulation is carried out on a square lattice
with side of $L=1024$ and for $T=2\times10^5$ time steps of
$\Delta t=0.001$. The surface height range is given in units of
lattice spacing set to unity.

\begin{figure}
\includegraphics[width=0.8\columnwidth]{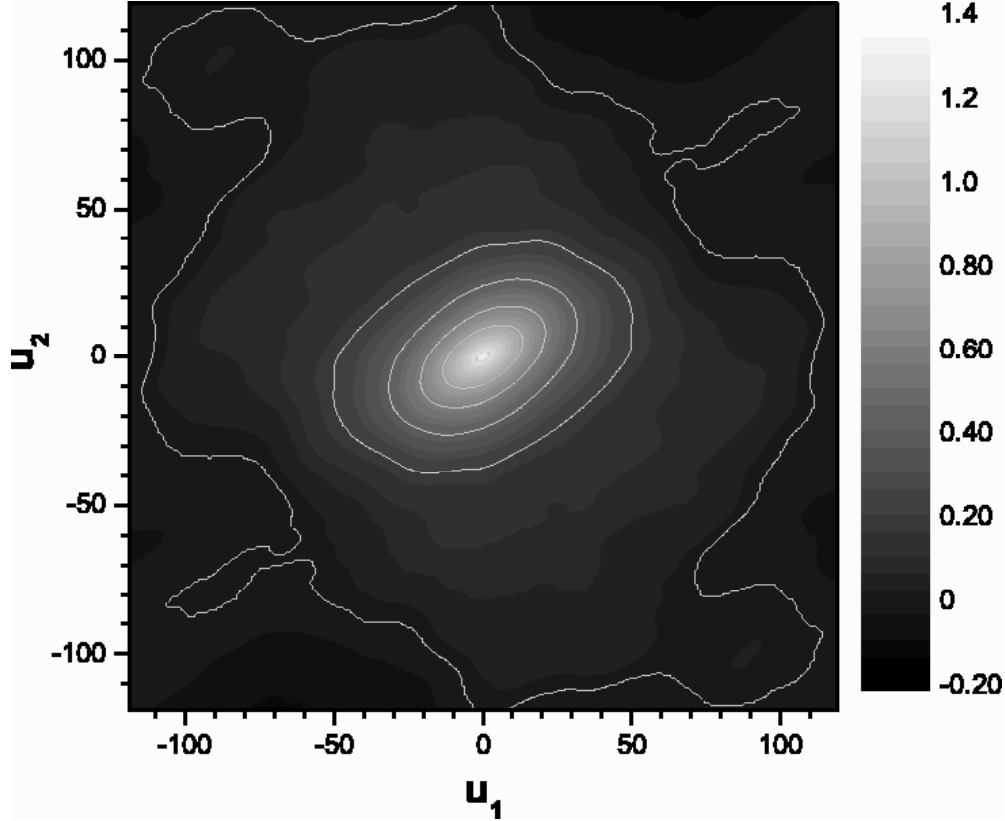}
\caption{\label{f:akpzacf30} Sample AcF for an image of the AKPZ
simulated morphology illustrated in Fig.~\ref{f:akpzsurf}, however
rotated on angle $30^{\circ}$. }
\end{figure}

In a typical experimental circumstances, the axes of anisotropy
are rarely known and need to be inferred and quantified from a
image of the morphology \cite{TRA99wear}. To reckon with this we
``record'' smaller, (512$\times$512), images, which are rotated at
angles $\psi_0=0^{\circ}$, $10^{\circ}$, $30^{\circ}$, and
$60^{\circ}$ with respect to the $x$-axis of the simulated
surface. The picture in Fig.~\ref{f:akpzsurf} represents the image
for $\psi_0=0^{\circ}$. The images for $\psi_0\neq 0^{\circ}$ are
obtained using a simple, based on the four nearest neighbor points
interpolation. Then for every image we compute the AcF estimate
(\ref{eq:S2dAcF}), an example of which for $\psi_0=30^{\circ}$ is
shown in Fig.~\ref{f:akpzacf30}. Taking a more systematic
approach, rather than the correlation length, we consider sections
of AcF defined by $aA(0)\leq A(\vec{u})\leq (a+\Delta{a})A(0)$ for
several levels $0<a<1$ and a fixed width of $\Delta a=0.04$. We
project the values of $A(\vec{u})$ within each section on the
$(u_1, u_2)$ plane and fit these points by an ellipse. The
direction of the axis of asymmetry and the aspect ratio is
evaluated from the parameters of these ellipses. In the actual
fits we have used four levels of $a$: 0.2, 0.4, 0.6, and 0.8.

\begin{figure}
\includegraphics[width=0.8\columnwidth]{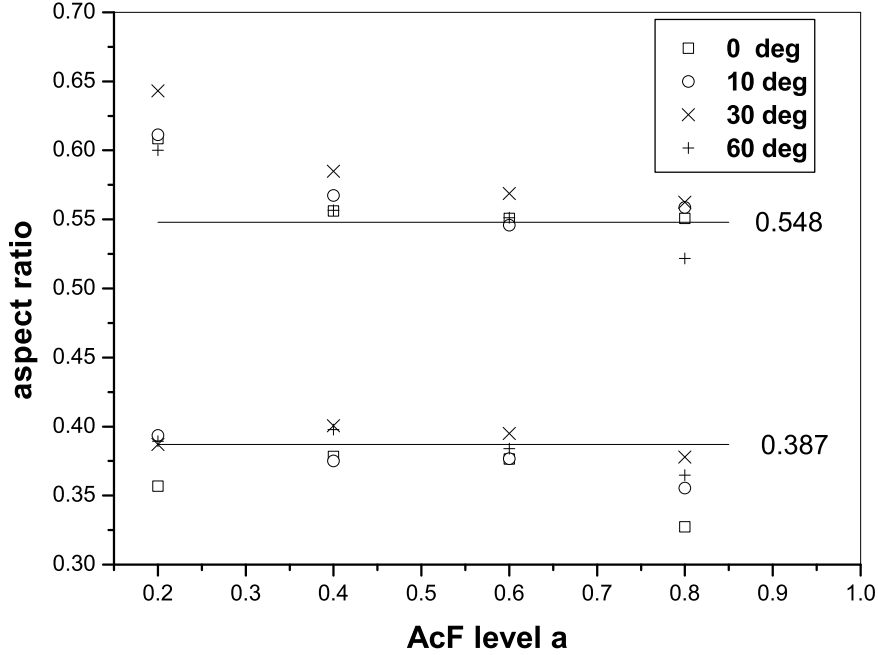}
\caption{\label{f:aspect} Retrieved aspect ratio $\gamma$ from
four AcF sections marked by levels $a$ and for four angles of
rotation of the recorded image, see the legends. Upper and bottom
data represent results from two simulated surfaces, see the text.
The horizontal lines indicate the expected values of the aspect
ratio, $\gamma=\sqrt{\nu_y/\nu_x}$. }
\end{figure}

The obtained results show a discrepancy typically  within $\pm
2^{\circ}$ from the expected direction of the anisotropy. In few
cases only, all associated with the lowest level of $a$,  the
discrepancy is up to $\pm 4^{\circ}$. More interesting are the
inferred values of the aspect ratio. These are plotted in
Fig.~\ref{f:aspect} for all four angles of rotation and AcF levels
$a$ and for two simulated surfaces, studied independently. The
parameters of the first simulation are the same as those used to
produce Fig.~\ref{f:akpzsurf}. The lattice size, the time step,
and total integration time for the second simulation are also the
same, however, with: $\nu_x=1$, $\nu_y=0.15$, $\lambda_x=10$,
$\lambda_x=1.4$, and $D=0.2$ ($r_{\nu}=r_{\lambda}=0.15$). The
values of $\gamma$ for the first simulation are grouped around
$0.548$, upper part of the figure, whereas for the second ---
around $0.387$ bottom part of the figure. Both these values
correspond to the respective $\sqrt{r_{\nu}}$ values used in the
simulations. To understand this, we rescale (\ref{eq:akpz0}) in a
manner different from the one used prior to numerical integration;
namely, $x = \sqrt{\nu_x}\widetilde{x}$, $y =
\sqrt{\nu_y}\widetilde{y}$, $h =
\left(\sqrt{2D}/\nu_x^{1/4}\nu_y^{1/4}\right)\widetilde{h}$, and
$\eta= (\sqrt{2D}/\nu_x^{1/4}\nu_y^{1/4})\widetilde{\eta}$,
arriving at

\begin{equation}\label{eq:akpz1}
   \partial_t\widetilde{h}=
   \partial_{\widetilde{x}}^2\widetilde{h}+
   \partial_{\widetilde{y}}^2\widetilde{h}+
   \frac{\varepsilon_x}{2}
   \left(\partial_{\widetilde{x}}\widetilde{h}\right)^2+
   \frac{\varepsilon_y}{2}
   \left(\partial_{\widetilde{y}}\widetilde{h}\right)^2 +
   \widetilde{\eta}.
\end{equation}

\noindent In the above equation,
$\varepsilon_x=\lambda_x\sqrt{2D}/\nu_x^{3/2}r_{\nu}^{1/4}$ and
$\varepsilon_y=\left(r_{\lambda}/r_{\nu}\right)\varepsilon_x$ and
hence at the fixed point of the RG, Eq.~(\ref{eq:akpz1}) is
equivalent to the isotopic KPZ equation. Therefore, the
anisotropic surface in this case is obtained by just rescaling the
isotropic surface in $x$ and $y$ directions by $l_x=\sqrt{\nu_x}$
and $l_y=\sqrt{\nu_y}$, respectively. The latter leads to aspect
ratio of $\gamma=\sqrt{r_{\nu}}$, which as demonstrated in
Fig.~\ref{f:aspect} is imprinted in the sample AcF,
Eq.~(\ref{eq:S2dAcF}). A somewhat larger discrepancy from
$\sqrt{r_{\nu}}$ observed at the lowest level fits can be
attributed to a greater relative variability of the SAcF. The
latter can be estimated crudely from the variance
(\ref{eq:S2dAcF-var}) by substituting the SAcF for the unknown
true AcF. For the two simulations used in Fig.~\ref{f:aspect}, we
obtain an increase from 12\% at $\vec{u}=0$, up to about 23\% for
points $\vec{u}$ at which $A(\vec{u})=0.2A(0)$. As a final remark,
the one-loop RG approximation of the AKPZ equation indicates that
if $\lambda_y\neq 0$ the surfaces in this class are characterized
by two characteristic lengths, $l_x$ and $l_y$, even when the
morphology has not yet evolved to the RG fixed point. In addition,
the characteristic length scale linearly, $l_y\sim l_x$,
Ref.~\cite{Wolf91prl}. This suggests that the approach undertaken
in this section may be suitable for characterization of more
generic AKPZ morphologies. Further numerical simulations, however,
are needed to confirm this assertion.

\section{Conclusions}

In this paper we have suggested an estimate for the autocovariance
function (AcF) of a homogenous random field in arbitrary dimension
$d$. The estimate, Eq.~(\ref{eq:SdAcF}), is constructed as to
represent the discrete and finite Fourier transform of the
periodogram estimate of the field's power spectrum; it is
identical to the standard AcF estimate in the first and the
$2^{d-1}+1$-th quadrants but differs in all other quadrants. As it
should be, the estimate is positive semidefinite. On the basis of
(\ref{eq:SdAcF}), we have suggested new estimates for the field's
structure function and power spectrum. We have also derived
expressions for the covariance, consequently for the variance of
the AcF estimate in two dimensions under the simplifying
assumption that the field is Gaussian and with a known mean.

Perhaps the most important advantage of the new sample AcF over
the standard estimate lays in the fact that it captures the
anisotropy of the field in all spatial directions. The latter is
demonstrated on two examples. The first involves the morphology of
the Grab nebula observed by the Hubble space telescope. For sake
of comparison we presented plots of the standard AcF as well. The
second example involves surfaces simulated by numerically solving
the anisotropic Kardar-Parisi-Zhang (AKPZ) equation and is
considered in more detail. In particular, we have focused on the
case $\nu_y/\nu_x=\lambda_y/\lambda_x$, i.e. when the system is at
a fixed point of the dynamic renormalization group approximation
for the AKPZ equation. In this case the characteristic lengths of
the morphology  are two and are determined by $l_x=\sqrt{\nu_x}$
and $l_y=\sqrt{\nu_y}$. Hence, the surface can be viewed as a
simple benchmark for testing statistical methods that account for
anisotropy. We have shown that one can retrieve both the direction
and the aspect ratio of the anisotropy reasonably well from the
estimate (\ref{eq:SdAcF}) in two dimensions. This has been done on
several sections of the AcF and on two independent realizations.

\begin{acknowledgments}
It is a pleasure to thank S. Zhekov for valuable comments and
suggestions, Tz. Georgiev for organizing a discussion of our work
at the Sofia astrophysics seminar and J. Hester for bringing
reference \cite{Hester96apj} to our attention. This study was
supported by the Bulgarian fund for science under grant F1203.
\end{acknowledgments}

\bibliography{AY06arsub}

\end{document}